\documentclass[a4paper,12pt,one column]{article}
\usepackage{graphicx}
\usepackage{colordvi}
\usepackage{amssymb}
\usepackage{latexsym}
\usepackage{hyperref}
\begin{document}

\title{Evolution of Primordial Black Hole Mass Spectrum in Brans-Dicke theory}
\author{D. Dwivedee$^{*}$, B. Nayak$^{\dag}$ and L. P. Singh$^{\ddag}$ \\
Department of Physics, Utkal University,
Bhubaneswar 751004, India. \\
$^{*}$debabrata@iopb.res.in \\ $^{\dag}$bibeka@iopb.res.in \\ $^{\ddag}$lambodar$\_$uu@yahoo.co.in \\}
\date{ }
\maketitle

\begin{abstract}
We investigate the evolution of primordial black hole mass spectrum by including both accretion of radiation and Hawking evaporation within Brans-Dicke cosmology in radiation, matter and vacuum-dominated eras. We also consider the effect of evaporation of primordial black holes on the expansion dynamics of the universe. The analytic solutions describing the energy density of the black holes in equilibrium with radiation are presented. We demonstrate that these solutions act as attractors for the system ensuring stability for both linear and nonlinear situations. We show, however, that inclusion of accretion of radiation delays the onset of this equilibrium  in all radiation, matter and vacuum-dominated eras. 
\end{abstract}

\section{Introduction}
The Brans-Dicke (BD) theory \cite{brans and dicke}, a scalar-tensor theory of gravity, is regarded as a viable alternative of Einstein's general theory of relativity (GTR). In the BD theory, gravitational constant G is replaced by a time-dependent scalar field $\phi$ which couples to gravity with a coupling parameter $\omega$. BD theory goes over to GTR in the limit $\omega \to \infty$. Interestingly, BD type models arise as low energy effective actions of several higher dimensional Kaluza-Klein and string theories \cite{asm1,asm2,asm3}. BD theory has been used to understand many cosmological phenomena such as inflation \cite{johri,la}, early and late time behaviour of the universe \cite{sahoo1,sahoo2}, cosmic acceleration and structure formation \cite{bermar} and the coincidence problem \cite{nayak} etc. 

The primordial black holes (PBHs) are the black holes which could be formed in the early universe in wide mass range through various mechanisms such as inflation \cite{cgl,kmz}, initial inhomogeneities \cite{carr,swh}, phase transition and critical phenomena in gravitational collapse \cite{khopol1,khopol2,khopol3,khopol4,jedam1,jedam2,jedam3}, bubble collision \cite{kss}, or the decay of cosmic loops \cite{polzem1,polzem2}. The formation masses of PBHs are so small that some of them are completely evaporated by the present epoch due to Hawking evaporation \cite{hawk}. PBHs which evaporate in early times could account for baryogenesis \cite{bckl,mds1,mds2} in the universe. On the other hand the longer lived PBHs could act as seeds for structure formation \cite{mor1,mor2,mor3,mor4,mor5} and could also form a significant component of dark matter \cite{blais1,blais2,blais3,blais4}. The most relevant issue concerning PBHs is their longivity which depends crucially on the effectiveness of various accretion processes.

In BD theory the possibility of black hole solutions was first proposed by Hawking \cite{hawking}. Using scalar-tensor gravity theories Barrow and Carr \cite{barrow and carr} studied PBH evaporation during various eras. It has been recently observed that in the context of Brans-Dicke theory and it's generalised version, inclusion of the effect of accretion of radiation leads to the prolongation of PBH lifetimes  \cite{nsmprd1,nsmprd2,mgs}. 


From the very beginning of the discovery of Hawking evaporation, it has been appreciated that the presence of a population of primordial black holes can have significant consequences for the evolution of the universe \cite{carr 1985}, which further motivates a discussion of the expected mass spectrum of primordial black holes. A number of works \cite{carr1,carr2,carr3,carr4,carr5,carr6,carr7} have been done in this regard for studying the evolution of a continuous mass spectrum of black holes. In this context, it is assumed by Carr \cite{carr 1975} that black holes exist with a power law mass spectrum. He also pointed out that formation of black holes over an extended mass range is possible only if the density perturbation spectrum in the universe has a power law Harrison-Zel'dovich form. A particular situation where primordial black holes form with a narrow range of masses was studied in the Reference  \cite{barrow 1991,barrow 1992}.
 
A detailed study on the evolution of population of primordial black holes in the very early universe was done by Barrow et al. \cite{paper}.  Where they included the effect of evaporation of PBHs on the expansion dynamics. We extend this analysis by including accretion of radiation by PBHs and taking Brans-Dicke theory as the theory of gravity in our present paper. We use a complete set of network equations which follow the evolution of a population of primordial black holes in an expanding universe containing either matter or radiation. The complete evolution equations consist of the evolution of the black hole spectrum, Friedmann equation describing the expansion of the space-time and an energy equation governing the evolution of the radiation energy density. 

\section{Evolution equations for the spectrum of black holes in Brans-Dicke theory}

Considering both accretion of radiation and Hawking evaporation, we obtain the equations which will determine the evolution of the spectrum of black holes and the effect of black holes and radiation on the evolution of the scale factor. To begin with,let us assume an isotropic Friedmann-Robertson-Walker space-time of zero curvature, i.e.
 \begin{equation}
ds^2=-dt^2 + a^{2}(t)(dx^2+dy^2+dz^2)
\end{equation}
where $a(t)$ represents the cosmic scale factor.

Assuming the number density of the initial black hole spectrum by a simple power-law form \cite{carr 1975}, we can write the initial number density of black holes between masses m and m+dm as
\begin{equation}
N(m,t=0)dm = A m^{-n} \Theta(m-km_{pl})dm, \hspace{0.5cm}  k\geq 1
\end{equation}
where $A$ represents the amplitude of the spectrum and $n$ represents the spectral index. We here introduce a minimum cut-off $km_{pl}$ through the theta function for the formation mass of the black holes in the spectrum where $k$ is an arbitrary dimensionless constant and $m_{pl}$ as the planck mass. The definition of theta ($\Theta $) function ensures that no black hole is formed with mass less than and equal to the cut-off mass. This initial cut-off helps to avoid the divergences at low masses. The above power law is assumed to hold for $n>2$ for which the total energy density does not diverge at large masses. As we take the accretion of radiation by black holes, the black holes take comparatively more time for their complete evaporation. During evaporation the mass of the black hole starts to decrease so the cut-off is reduced to make the number density non- zero. This cut-off vanishes when the initially lightest holes evaporate completely. From this initial spectrum we can determine the spectrum at all later times. Then at a given time(t), the number density of black holes with cut-off in the spectrum will be
\begin{equation}
N(t)=\int_{0}^{\infty} N(m,t)dm
\end{equation}
Hence the energy density of the black holes becomes
\begin{equation}
\rho_{BH}(t)= \int_{0}^{\infty}N(m,t)m dm 
\label{4}
\end{equation}
where the integrals are carried out at constant time.

In order to derive analytical results we assume all black holes form at the same time. In practice, a hole of mass $m_{BH}$ can not form until the horizon mass exceeds $m_{BH}$, implying that larger holes form later and hence there is a delay before the evaporation. The lightest black holes may evaporate in a time less than their formation time where this assumption can not be valid. But Barrow et.al. \cite{paper} have shown that this assumption is valid nearly beyond $10^{-37}s $. So our approximation breaks down only at extremely early times. 

The black hole number spectrum varies with time due to dilution by the expansion of the universe and the evaporation of the holes into radiation by Hawking process. Because of the Hawking evaporation the rate of decrease of the mass $m_{BH}$ of a single black hole is given by
\begin{equation}
\dot{m}_{BH} = -4\pi r_{BH}^2a_{H}T_{BH}^4
\label{5}
\end{equation}
where $r_{BH}=2Gm_{BH}$ represents the radius of the black hole, $a_{H}$ represents the Stefan-Boltzmann constant multiplied with number of degree of freedom available for radiation, $T_{BH}=\frac{1}{8\pi G m_{BH}}$ is the Hawking Temperature and $G$ is a time dependent quantity in Brans-Dicke theory. Using these expressions equation (\ref{5}) becomes
\begin{equation}
\dot{m}_{BH} = -\frac{a_{H}}{256\pi ^3} \frac{1}{G^2 m_{BH}^2}
\label{6}
\end{equation}
In Brans-Dicke theory, Barrow and Carr \cite{accpaper} have obtained the following solutions for $G$ for different eras, as
\begin{eqnarray}
G(t)= \left\{
\begin{array}{rr}
G_0\Big(\frac{t_0}{t_e}\Big)^{n_1} & (t<t_e)\\
G_0\Big(\frac{t_0}{t}\Big)^{n_1} & (t>t_e)
\end{array}
\right.
\label{7}
\end{eqnarray}   
where $t_{e}$ represents the era of radiation-matter equality, $t_{0}$ represents the present time, $G_{0}$ represents the present value of $G \simeq t_{pl}/m_{pl} $ and $ n_{1}$ is a parameter related to $\omega $ as $n_{1}=\frac{2}{4+3\omega}$.
Since solar system observations \cite{nature}  require that $\omega $ be large ($\omega $ $\geq $ $ 10^{4}$), $n_{1}$ is very small $(n_{1}\leq 0.00007)$.
         
Using the values of $G(t)$ from the equation (\ref{7}) we can write equation (\ref{6}) for radiation-dominated era as
\begin{equation}
\dot{m}_{BH} = -\frac{a_{H}}{256\pi ^3} \frac{1}{G_{0}^2 (\frac{t_{0}} {t_{e}})^{2n_{1}}m_{BH}^2}
\label{8}
\end{equation}
As a black hole accretes in radiation dominated era, its mass increases from its initial value $(m_{0})$ at the time $t=0$ to a maximum value $m_{c}$ at the time $t = t_{c}$ \cite{nsmprd1,nsmprd2}, where
\begin{equation}
m_{c}= \frac{m_{0}}{(1-\frac{3}{2} f)}
\end{equation}
and
\begin{equation}
t_{c}=\Big[{\frac{f}{{\frac{2}{3}} G_{0}^{-1} {\frac{a_{H}}{256 \pi ^3 G_{0}^2}} (\frac{t_{e}}{t_{0}})^{3n_{1}}}}\Big]^{1/2}m_c^2
\end{equation}
which demands that $f<2/3$.
Till the time $t_{c}$ accretion by black hole remains dominant over evaporation and after that evaporation becomes dominant.
Therefore, considering the evaporation alone from $t_{c}$ onwards the mass of the black hole reduces from $m_{c}$ to some other value m at any time t. Now we can solve equation (\ref{8}) by integration with proper limits and get
\begin{equation}
{m_{BH}^3}(t) = m_c^3 - {\frac{3a_{H}}{256\pi ^3}}{\frac{1}{{G_{0}^2}(\frac{t_{0}}{t_{e}})^{2n_{1}}}} (t-t_c)
\label{11}
\end{equation}
The time at which the black hole is completely evaporated is now found from equation (\ref{11}) as
\begin{equation}
t_{evap} = {\frac{256\pi ^3 }{3a_{H}}}G_0^2\Big(\frac{t_{0}}{t_{e}}\Big)^{2n_{1}}m_c^3 + t_c
\end{equation}

We are, however, interested to evaluate the spectrum at time $t$ by taking accretion and evaporation into account. Then the black holes having masses between $m_0$ and $m_0$+$dm_0$ evolve to masses $m$ and $m+dm$ in a time $t$, where the number density at time $t$ remains same to the original number density at time $t=0$. Now using the Jacobian factor we obtain the number density of the black hole spectrum at time $t$ in radiation dominated era as


\begin{eqnarray}
N(m,t)dm = A \Big(1-\frac{3}{2}f\Big)^{-n+1} m^{-n} \Big(1+{\frac{3a_H}{256\pi^3}}{\frac{1}{G_0^2(\frac{t_0}{t_e})^{2n_1}}}m^{-3}(t-t_c)\Big)^{-\frac{n+2}{3}}
 \nonumber \\ \Theta\Big[m-{\frac{km_{pl}}{(1-\frac{3}{2}f)}}\Big{\{}1-{\frac{3a_H}{256\pi^3}}{\frac{1}{G_{0}^2(\frac{t_0}{t_e})^{2n_1}}}\Big(1-\frac{3}{2}f\Big)^3 k^{-3}m_{pl}^{-3}(t-t_{c})\Big{\}}^{\frac{1}{3}}\Big{]}dm
\label{13}
\end{eqnarray}
Equation (\ref{13}) tracks the behaviour of the cut-off which at later time takes the form 
\begin{equation}
m_{cut-off}(t)=\frac{km_{pl}}{(1-\frac{3}{2}f)}\Big{\{}1-{\frac{3a_H}{256\pi^3}}
{\frac{1}{G_0^2(\frac{t_0}{t_{e}})^{2n_1}}} \Big(1-\frac{3}{2}f\Big)^3k^{-3}m_{pl}^{-3}(t-t_c)\Big{\}}^\frac{1}{3}
\label{14}
\end{equation}
This cut-off found through the theta $(\Theta )$ function shows that as long as the accretion is dominant ($t<t_c$) the cut-off value increases and beyond $t_c$ evaporation becomes dominant making the cut-off value monotonically decrease. The cut-off value reaches zero at the time
\begin{equation}
t = \Big{(}\frac{3a_H}{256\pi^3}\Big{)}^{-1} G_0^2 \Big(\frac{t_0}{t_e}\Big)^{2n_1}k^3m_{pl}^3\Big(1-{\frac{3}{2}f}\Big)^{-3} +t_c 
\label{15}
\end{equation} 
and vanishes from the picture after that time.
The number density of black hole spectrum decreases at late times when the cut-off vanishes due to the complete evaporation of lightest blackholes.
Substituting equation (\ref{13}) in equation (\ref{4}) the black hole energy density becomes
\begin{eqnarray}
\rho_{BH}(t) =\int_{0}^{\infty} A \Big(1-\frac{3}{2}f\Big)^{-n+1} m^{-n+1} \Big(1 + {\frac{3a_H}{256\pi^3}}{\frac{1}{G_0^2(\frac{t_0}{t_e})^{2n_1}}}m^{-3}(t-t_c) \Big)^{-\frac{n+2}{3}}
 \nonumber \\ \Theta\Big[m-{\frac{km_{pl}}{(1-\frac{3}{2}f)}}\Big{\{}1 - {\frac{3a_H}{256\pi^3}}{\frac{1}{G_{0}^2(\frac{t_0}{t_e})^{2n_1}}}\Big(1-\frac{3}{2}f\Big)^3 k^{-3}m_{pl}^{-3}(t-t_{c})\Big{\}}^{\frac{1}{3}}\Big]dm
\label{16}
\end{eqnarray}


The amount of energy density transferred by the PBHs through evaporation between times t and t+dt is given by 
\begin{equation}
dE=\rho_{BH}(t)-\rho_{BH}(t+dt)=-\frac{\partial \rho_{BH}}{\partial t} dt
\label{17}
\end{equation}
As the cut-off position varies with time we can compute the rate of the energy density transfer, by using equation (\ref{16}) as
\begin{eqnarray}
\frac{dE}{dt} = A\Big(\frac{n+2}{3}\Big)\frac{3a_H}{256\pi^3}{\frac{1}{{G_0}^2(\frac{t_0}{t_e})^{2n_1}}}\Big(1-\frac{3}{2}f\Big)^{-n+1}\int_{m_{cut-off}(t)}^{\infty}m^{-n-2}\Big{[}1 +\frac{3a_H}{256\pi^3} \nonumber \\ \frac{1}{G_0^2(\frac{t_0}{t_e})^{2n_1}} m^{-3}(t-t_c)\Big{]}^{-(\frac{n+5}{3})}dm -\frac{A}{3} k^{-n-1}m_{pl}^{-n-1}\frac{3a_H}{256\pi^3}\frac{1}{{G_0^2}(\frac{t_0}{t_e})^{2n_1}}(1-\frac{3}{2}f)^{2}\nonumber \\ \Big[1-\frac{3a_H}{256\pi^3} \frac{1}{G_0^2(\frac{t_0}{t_e})^{2n_1}}k^{-3}m_{pl}^{-3} \Big(1-\frac{3}{2}f\Big)^3(t-t_c)\Big]^\frac{1}{3} \Theta\Big[\frac{k^3m_{pl}^3(1-\frac{3}{2}f)^{-3}}{\frac{3a_H}{256\pi^3}\frac{1}{G_0^2(\frac{t_0}{t_e})^{2n_1}}}+t_c-t\Big]
\label{18}
\end{eqnarray}
where
\begin{equation}
m_{cut-off}(t)=max\Big[0, \frac{km_{pl}}{(1-\frac{3}{2}f)}\Big{\{}1-\frac{3a_H}{256\pi^3}\frac{1}{G_0^2(\frac{t_0}{t_e})^{2n_1}} \Big(1-\frac{3}{2}f\Big)^3k^{-3}m_{pl}^{-3}(t-t_c)\Big{\}}^\frac{1}{3}\Big]
\label{19}
\end{equation}
From equation (\ref{18}) it is clear that the first term is the only contributor at late times. As the cut-off disappears at sufficiently late times the second term vanishes due to the presence of $\Theta$ - function in it.

Now invoking expansion of the universe, as the volume of the universe increases the black hole density starts to dilute. Hence the spectral amplitude $A$ will be replaced by A${\alpha}^{-3}$, where ${\alpha}(t)=\frac{a(t)}{a(0)}$. From this definition we have $\alpha(0)=1$ at the formation time of black holes $(t=0)$.
Now, the equation of motion containing Friedmann equation for describing the expansion of space-time and energy equation governing the evolution of the radiation energy density $\rho_R$ in radiation dominated era in Brans-Dicke theory can be written as,
\begin{equation}
\Big(\frac{\dot\alpha}{\alpha}\Big)^2=\frac{8\pi}{3}G_0 \Big(\frac{t_0}{t_e}\Big)^{n_1}(\rho_R+\rho_{BH})
\label{20}
\end{equation} 
\begin{equation}
\dot\rho_{R}=-4\frac{\dot\alpha}{\alpha}\rho_R +\frac{dE}{dt}
\label{21}
\end{equation}
In order to solve equations (\ref{20}) and (\ref{21}) we have to assume sufficiently late times so that the lightest black holes are completely evaporated making the cut-off vanish from the system. This corresponds to the time $ t>\Big[\Big(\frac{3a_H}{256\pi^3}\Big)^{-1}G_0^2\Big(\frac{t_0}{t_e}\Big)^{2n_1}k^3m_{pl}^3(1-\frac{3}{2}f)^{-3}+t_c\Big]$.
Implementing this condition we can evaluate $\rho_{BH}$ and $\frac{dE}{dt}$ as 
\begin{equation}
\rho_{BH}(t)=\rho_{BH0}\Big(1-\frac{3}{2}f\Big)^{-n+1}\alpha^{-3}(t)\Big(t-t_c\Big)^\frac{2-n}{3},\hspace{0.5cm} n>2
\label{23}
\end{equation}
and
\begin{equation}
\frac{dE}{dt}=E_0\Big(1-\frac{3}{2}f\Big)^{-n+1}\alpha^{-3}(t)\Big(t-t_c\Big)^{\frac{-1-n}{3}},\hspace{0.5cm} n>2
\label{24}
\end{equation}
where $\rho_{BH_0}$ and $E_0$ are constants given by
\begin{equation}
\rho_{BH0}=-\frac{A}{3}\Big(\frac{3a_H}{256\pi^3}\frac{1}{G_0^2(\frac{t_0}{t_e})^{2n_1}}\Big)^{\frac{2-n}{3}} B\Big(\frac{n-2}{3}, \frac{4}{3} \Big)
\label{25}
\end{equation}
\begin{equation}
E_0=-\frac{A}{3}{\Big(\frac{n+2}{3}\Big)}\Big(\frac{3a_H}{256\pi^3}\frac{1}{G_0^2(\frac{t_0}{t_e})^{2n_1}} \Big)^{\frac{2-n}{3}} B\Big(\frac{n+1}{3},\frac{4}{3}\Big)
\label{26}
\end{equation}
where $B\Big(\frac{n-2}{3},\frac{4}{3}\Big)$ and $B\Big(\frac{n+1}{3},\frac{4}{3}\Big)$ are the Beta functions.\\
From equations (\ref{23})and (\ref{24}) we find

\begin{equation}
\frac{dE}{dt}=\frac{(n-2)}{3}\frac{\rho_{BH}}{(t-t_c)}
\label{27}
\end{equation}
Using equations (\ref{23}) and (\ref{27}) in equations (\ref{20}) and (\ref{21}) the evolution equations of the universe in radiation dominated era can be written in a simpler form as
\begin{equation}
{\dot\alpha}^2(t)=\frac{8\pi}{3}G_0\Big(\frac{t_0}{t_e}\Big)^{n_1}{\alpha}^2(t)\Big[\rho_R(t)+\rho_{BH0}\Big(1-\frac{3}{2}f\Big)^{-n+1}\alpha^{-3}(t)\Big(t-t_c\Big)^{\frac{2-n}{3}}\Big]
\label{28}
\end{equation}
 
\begin{equation}
{\dot\rho_R}(t)=-4\frac{\dot\alpha(t)}{\alpha(t)}\rho_R(t)+\frac{n-2}{3}\rho_{BH0}\Big(1-\frac{3}{2}f\Big)^{-n+1}{\alpha^{-3}}(t)\Big(t-t_c\Big)^\frac{-1-n}{3}
\label{29}
\end{equation}

 It is clear from equations (\ref{23}), (\ref{24}), (\ref{28}) and (\ref{29}) that in the limit $f \to 0$ and $t_c=0$, these equations take the form as obtained in Barrow et al. \cite{paper}. It may be noted that non-zero $f$ and $t_c$ capture the effect of accretion of radiation by the PBHs in radiation-dominated era. From equation (\ref{23}), we find that  with increase in both accretion efficiency ($f$) and upper limit of accretion time ($t_c$), black hole density increases. Similar observations apply for the rate of energy transferred by PBHs through evaporation, which is concluded from equation (\ref{24}). 

\section {Black holes in a radiation-dominated era}

\subsection{An exact equilibrium solution}

We begin by considering a power-law solution of equations (\ref{28}) and (\ref{29}) for $(t>0)$ as
\begin{equation}
\alpha(t)=\alpha_0t^r,\hspace{1.5cm}     \rho_R(t)=\rho_{R0}t^s
\label{30}
\end{equation} 
By the use of Equation (\ref{30}), we get from equations (\ref{28}) and (\ref{29}) that
\begin{equation}
s=-2,\hspace{1.5cm} r=\frac{8-n}{9}
\label{31}
\end{equation}
where we have taken $\Big(1-\frac{t_c}{t} \Big)$ as constant which strictly holds for $t\gg t_c$.\\
The coefficients of power laws $\alpha_0$ and $\rho_{R0}$ can be found by using equations (\ref{30}) and (\ref{31}) into the evolution equations (\ref{28}) and (\ref{29}) as
\begin{equation}
\rho_{R0}=\frac{3}{8\pi G_0} \Big(\frac{8-n}{9}\Big)^2 \Big(\frac{t_e}{t_0}\Big)^{n_1}\Big[1+\frac{2(7-2n)}{3(n-2)}\Big(1-\frac{t_c}{t}\Big)\Big]^{-1}
\label{32}
\end{equation}
\begin{eqnarray}
\rho_{BH0}\alpha_0^{-3}=\frac{1}{324 \pi G_0} \frac{(8-n)^2(7-2n)}{(n-2)} \Big(\frac{t_0}{t_e}\Big)^{-n_1}\Big(1-\frac{3}{2}f\Big)^{n-1} \Big(1-\frac{t_c}{t}\Big)^{(n+1)/3} \nonumber \\ \Big[1+\frac{2(7-2n)}{3(n-2)}\Big(1-\frac{t_c}{t}\Big)\Big]^{-1}
\label{33}
\end{eqnarray}
where $\alpha_0$ is determined in terms of the known $\rho_{BH0}$ which is given by equation (\ref{25}). These values of coefficients imply that the above power-law solutions exist only for $n\in$ (2, 7/2). With this range of $n$, equation (\ref{31}) implies $r\in $ (1/2, 2/3). Now equation (\ref{30}) confirms that the expansion power law is  intermediate between that of pure radiation-domination and nearly matter-domination considering negligible value of $n_1$. Hence the presence of black holes results in a different expansion rate compared with that of a conventional Friedmann cosmology. 

Now the ratio of energy densities for this solution can be calculated as
\begin{equation}
\frac{\rho_{BH}(t)}{\rho_R(t)}=\frac{2}{3}\Big(\frac{7-2n}{n-2}\Big) \Big(1-\frac{t_c}{t}\Big)
\label{34}
\end{equation}
This equation shows that the ratio depends on spectral index $n$ and $\frac{t_c}{t}$. But at sufficient late time, $\frac{t_c}{t} \to 0$ and the ratio becomes a fixed constant depending only on the spectral index $n$. So the solution represents an equilibrium between black holes and radiation at a late time $t$ decided by time $t_c$ beyond which evaporation dominates over accretion of radiation. The late time $(t>t_c)$ equilibrium exists because the decrease in radiation density due to the expansion of space-time relative to the black holes is compensated by the radiation energy evaporated from the black holes. 

At sufficiently late time the above ratio gives that as $ n\to$ 2 from above the  black hole density dominates and correspondingly $ r\to2/3$ giving  nearly matter-dominated state as $n_1$ is very small, whereas  $ n\to 7/2$ from below radiation density dominates and correspondingly $ r\to 1/2$ gives pure radiation-dominated state. This is practically the same result as obtained by Barrow et al. \cite{paper} though in our case it is true only for $t\gg t_c$ due to consideration of accretion of radiation.

\subsection{Stability of the equilibrium solution}
It is pointed out by Barrow etal. \cite{paper} that the above power-law solution is a particular one for its fixed coefficients and fixed powers of $t$. Such a solution will not be valid for arbitrary initial conditions. So they modified the above analysis by using a co-ordinate transformation to make the above solution an attractor for a wide class of initial conditions. 
Now we undertake similar analysis in BD theory so that both linear and non-linear stability of the exact solution can be studied.

We use equations (\ref{32}) and (\ref{33}) in equation (\ref{30}) to set

\begin{equation}
\alpha(t)=\rho_{BH0}^{\frac{1}{3}}\xi(t)t^{\frac{8-n}{9}}
\label{35}
\end{equation}

\begin{equation}
\rho_R\alpha^4=\rho_{BH0}^{\frac{4}{3}}\eta(t)t^{\frac{14-4n}{9}}
\label{36}
\end{equation}
where $\xi$ and $\eta$ are two new dynamical variables. Here we replace time $t$ by a new time co-ordinate $\tau$ such that $t=e^{\tau}$ and  we indicate  all the derivatives with respect to $\tau $ by primes.
Using these equations we can write the evolution equations (\ref{28}) and (\ref{29}) in terms of the new variables $\xi$ and $\eta $ as 
\begin{eqnarray}
\xi^{'}=\sqrt{\frac{8\pi}{3}G_0 \Big(\frac{t_0}{t_e}\Big)^{n_1}\Big[\eta\xi^{-2}+\Big(1-\frac{3}{2}f\Big)^{-n+1}\xi^{-1}\Big(1-\frac{t_c}{t}\Big)^{\Big(\frac{2-n}{3}\Big)}\Big]}-\Big(\frac{8-n}{9}\Big)\xi & ,
\label{37}
\end{eqnarray}
\begin{equation}
\eta^{'}=-\Big(\frac{14-4n}{9}\Big)\eta+\Big(\frac{n-2}{3} \Big)\Big(1-\frac{3}{2}f \Big)^{-n+1} \Big(1-\frac{t_c}{t}\Big)^{\frac{-n-1}{3}}\xi
\label{38}
\end{equation}
In these new coordinates, the exact solutions of the above two equations corresponds to a critical point which is found in the positive quadrant of the phase plane as
\begin{eqnarray}
\xi_0=\sqrt[3]{\frac{8\pi G_0}{3} \Big(\frac{9}{8-n}\Big)^2\Big(\frac{t_0}{t_e}\Big)^{n_1}\Big(1-\frac{3}{2}f\Big)^{-n+1} \Big(1-\frac{t_c}{t}\Big)^{\frac{-n-1}{3}} \Big(\frac{8-n}{14-4n}-\frac{t_c}{t}\Big)}
\label{39}
\end{eqnarray}
\begin{eqnarray}
\eta_0=\frac{3(n-2)}{2(7-2n)}\Big(1-\frac{3}{2}f\Big)^{-n+1} \Big(1-\frac{t_c}{t}\Big)^{\frac{-n-1}{3}} ~~~~~~~~~~~~~~~~~~~~~~~~~~~~~~~~~~~~ \nonumber \\ \sqrt[3]{\frac{8\pi G_0}{3} \Big(\frac{9}{8-n}\Big)^2\Big(\frac{t_0}{t_e}\Big)^{n_1}\Big(1-\frac{3}{2}f\Big)^{-n+1} \Big(1-\frac{t_c}{t}\Big)^{\frac{-n-1}{3}} \Big(\frac{8-n}{14-4n}-\frac{t_c}{t}\Big)}
\label{40}
\end{eqnarray}

Above values of $\xi_0$ and $\eta_0$ imply that accretion plays a role in fixing the critical point. For increasing $f$ subject to condition $f<2/3$ values, $\xi_0$ and $\eta_0$ increase. Also accretion delays the approach to the critical point. But for a particular value of $f$ and at sufficient late time, $\xi_0$ and $\eta_0$ become constant and act as a true critical point.

The Sketch of the phase plane for the system (for $f=0$ and large $t$) is shown in figure-1. From this figure, we find that all initial conditions tend towards the exact solution at late time.

\begin{figure}[h]
\centering
\includegraphics[scale=0.4]{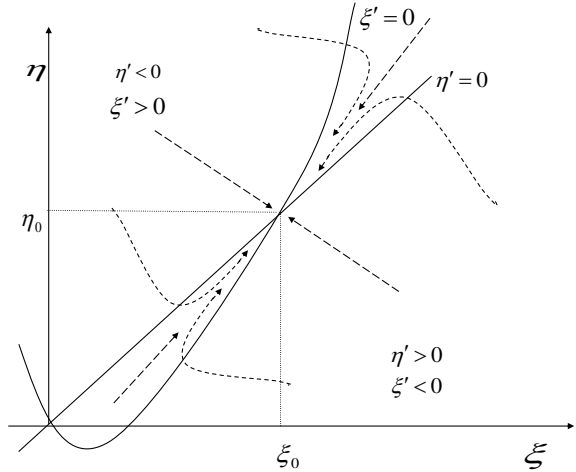}
\caption{The phase plane for black holes in equilibrium with radiation in an expanding universe. Our analytic solution corresponds to the critical point ($\xi_0$, $\eta_0$) given by equations (\ref{39}) and (\ref{40}) and is seen to be an attractor for all initial conditions in the first quadrant (the physical region).}
\label{fig1}
\end{figure}
Now we study the linear and non-linear stability of the critical point by following analysis of Simmons \cite{simmon}. First we translate the co-ordinates so as to bring the critical point to the origin. Thus our new variables are defined as $\rho=\xi- \xi_0$ and $\sigma=\eta -\eta_0 $.\\
Linearizing with respect to these variables we obtain the linear stability equations as
\begin{equation}
\rho^{'}=-\Big(\frac{1+n}{9}\Big)\rho + \Big(\frac{7-2n}{9} \Big)\Big(1-\frac{3}{2}f \Big)^{n-1} \Big(1-\frac{t_c}{t}\Big)^{(n-2)/3} \sigma
\label{41}
\end{equation}

\begin{equation}
\sigma^{'}=-\Big(\frac{14-4n}{9} \Big)\sigma +\Big(\frac{n-2}{3}\Big)\Big(1-\frac{3}{2}f \Big)^{-n+1}  \Big(1-\frac{t_c}{t}\Big)^{(-n-1)/3} \rho
\label{42}
\end{equation}
From these, we obtain the auxiliary equation as \cite{simmon}
\begin{equation}
p^2+\Big(\frac{5-n}{3} \Big)p+\frac{1}{81}\Big[(1+n)(14-4n)-3(n-2)(7-2n) \Big(1-\frac{t_c}{t}\Big)^{-1}\Big]=0
\label{mudhi}
\end{equation}
The roots of this auxiliary equation is given by
\begin{equation}
p=\frac{1}{2}\Big[-\Big(\frac{5-n}{3}\Big)\pm \sqrt{\Big(\frac{5-n}{3}\Big)^2-\frac{4}{81}\Big{\{}(1+n)(14-4n)-3(n-2)(7-2n) \Big(1-\frac{t_c}{t}\Big)^{-1}\Big{\}}}\Big]
\label{43}
\end{equation} 
But the roots of the auxiliary equation actually determine the stability of the critical point. Since the roots are both real, distinct and negative, the critical point is asymptotically stable and the same sign (i.e. -ve) implies that it is a node i.e all trajectories enter the critical point.

It may be mentioned here that we had neglected the non-linear terms in setting up the linear set of equations. This stands aposteriori justified as long as one works near critical points. Thus the stability shown above for linear system of equations also implies stability for full non-linear system. Thus our exact solution is asymptotically stable and is an attractor for all physical initial conditions. 

\subsection{The approach to the attractor}

 Now we estimate the rate of approach to the attractor by using the assumption $\rho_{BH} \ll \rho_R$ in equations (\ref{28}) and (\ref{29}). This assumption actually gives a pure radiation-dominated solution as
\begin{equation}
\alpha(t)=\Big{\{}\frac{32\pi \rho_{R0}G_0}{3}\Big(\frac{t_0}{t_e} \Big)^{n_1} \Big{\}}^{1/4}\Big[t+\sqrt{\frac{3G_0^{-1}}{32\pi \rho_{R0}}\Big(\frac{t_e}{t_0}\Big)^{n_1}} \Big]^{1/2}
\label{44}
\end{equation}
 
\begin{equation}
\rho_R(t)=\Big{\{}\frac{32\pi G_0}{3}\Big(\frac{t_0}{t_e} \Big)^{n_1} \Big{\}}^{-1}\Big[t+\sqrt{\frac{3G_0^{-1}}{32\pi \rho_{R0}}\Big(\frac{t_e}{t_0} \Big)^{n_1}} \Big]^{-2}
\label{45}
\end{equation}
Here $\rho_{R_0}$ is the radiation density at the formation time of black hole (i.e. $t=0$).
The above equation (\ref{45}) shows that the radiation density falls off as $t^{-2}$.

Again using equation (\ref{44}) in equation (\ref{23}), we can determine the time variation of black hole density as 
\begin{eqnarray}
\rho_{BH}=\rho_{BH0}\alpha^{-3}(t-t_c)^{(2-n)/3}  \Big(1-\frac{3}{2}f\Big)^{-n+1} \sim t^{-(2n+5)/6} \Big(1-\frac{3}{2}f\Big)^{-n+1} \Big(1-\frac{t_c}{t}\Big)^{(2-n)/3}
\label{46}
\end{eqnarray}
Equation (\ref{46}) tells that with increase in the value of accretion efficiency ($f$) black hole energy density increases. But for a constant accretion efficiency ($f$), black hole density is controlled by spectral index ($n$). For $n<7/2$, where actually attractor solution exists, black hole density falls at a slower rate than radiation density. Since initial radiation density is much larger than initial black hole density, the domination of black hole density is possible at a sufficient large time. Closer the value of $n$ to 7/2, more time is required for black hole density term to become important in Friedmann equation.

It is also simple to show that as long as $\rho_{BH} \ll \rho_R$ the energy fed into radiation from the holes will never be significant, so there is no possibility that the evaporation can help to maintain radiation domination.

The approach to the attractor is gauged by the evolution of the quantity
\begin{eqnarray}
\frac{\rho_{BH}}{\rho_R}\sim t^{(7-2n)/6} \Big(1-\frac{3}{2}f\Big)^{-n+1} \Big(1-\frac{t_c}{t}\Big)^{(2-n)/3}
\label{47} 
\end{eqnarray}
This ratio clearly depends on accretion; with increase in the value of accretion efficiency $f$ and $t_c$, it increases.


\section{Evolution of blackholes in a matter-dominated era}

We now study evolution of black hole spectrum in a matter-dominated universe.
For this we have to generalise our equations of motion to include matter terms
 as well as radiation terms. Let us introduce an additional parameter $\beta$ to measure the fraction of the evaporated energy 
of the black holes which goes to the radiation and hence (1-$\beta$) is the 
fraction going into matter (i.e. into particles which become rapidly non-relativistic), where $\beta$ could in general be a function of the mass  
distribution of the black holes but here it is assumed to be a constant.  

Now the Friedmann equation and energy equations
are given by
\begin{eqnarray}
\dot\alpha^2=\frac{8\pi}{3}G_0\Big(\frac{t_0}{t}\Big)^{n_1}\alpha^2\Big[\rho_M+\rho_R+\rho_{BH0}\Big(1-\frac{3}{2}f\Big)^{-n+1}\alpha^{-3}(t-t_c)^{\frac{2-n}{3}}\Big] \nonumber \\ - \frac{n_1}{t}\dot\alpha\alpha+\frac{\omega}{6}\Big(\frac{n_1}{t}\Big)^2\alpha^2
\label{48}
\end{eqnarray}

\begin{equation}
\dot\rho_R=-4\frac{\dot\alpha}{\alpha}\rho_R+\beta\Big(\frac{n-2}{3}\Big)\rho_{BH0}\Big(1-\frac{3}{2}f\Big)^{-n+1}\alpha^{-3}(t-t_c)^{\frac{-1-n}{3}}
\label{49}
\end{equation}

\begin{equation}
\dot\rho_M=-3\frac{\dot\alpha}{\alpha}\rho_M+(1-\beta)\Big(\frac{n-2}{3}\Big)\rho_{BH0}\Big(1-\frac{3}{2}f\Big)^{-n+1}\alpha^{-3}(t-t_c)^{\frac{-1-n}{3}}
\label{50}
\end{equation}
where $\rho_M$ represents the matter energy density. Here we have used the assumption that  we are past the cut-off. 
Assuming $\rho_R$ and $\rho_{BH}$ terms are to be negligible in comparision to $\rho_M$ term in matter-dominated universe we can drop them from the above equations (\ref{48}) and (\ref{50}) to get the standard matter dominated solutions to these equations as
\begin{equation}
\alpha(t)=C\Big[\Big(\frac{8\pi}{3}G_0t_0^{n_1}\Big)^{1/3}\Big{\{}\Big(\frac{2-n_1}{3}\Big)^2 +\Big(\frac{2-n_1}{3}\Big)n_1-\frac{\omega}{6}n_1^2\Big{\}}^{-1/3}\Big]t^{\frac{2-n_1}{3}}
\label{51}
\end{equation}
 
\begin{equation}
\rho_M(t)=\Big(\frac{8\pi}{3}G_0t_0^{n_1}\Big)^{-1}\Big{\{}\Big(\frac{2-n_1}{3}\Big)^2 + \Big(\frac{2-n_1}{3}\Big)n_1-\frac{\omega}{6}n_1^2\Big{\}}t^{n_1-2}
\label{52}
\end{equation}
where C is a constant determined from the initial conditions.
Now using the above scale factor we can obtain an expression for black hole energy density which is given in previous section (equation (\ref{23})) as
\begin{equation}
\rho_{BH}(t)=\rho_{BH0}\Big(1-\frac{3}{2}f\Big)^{-n+1}\alpha_0^{-3}\Big(1-\frac{t_c}{t}\Big)^{\frac{2-n}{3}}t^{\frac{-4-n+3n_1}{3}}
\label{53}
\end{equation}
where $\rho_{BH0}$ is same as in the previous section and 
\begin{equation}
\alpha_0=C\Big(\frac{8\pi}{3}G_0t_0^{n_1}\Big)^{1/3}\Big{\{}\Big(\frac{2-n_1}{3}\Big)^2 +\Big(\frac{2-n_1}{3}\Big)n_1-\frac{\omega}{6}n_1^2\Big{\}}^{-1/3}
\label{54}
\end{equation}
Now the radiation equation given by equation (\ref{49}) becomes
\begin{equation}
\dot\rho_R=\Big(\frac{4n_1-8}{3t}\Big)\rho_R+\beta\Big(\frac{n-2}{3}\Big)\rho_{BH0}\Big(1-\frac{3}{2}f\Big)^{-n+1}\alpha_0^{-3}\Big(1-\frac{t_c}{t}\Big)^{\frac{-1-n}{3}}t^{\frac{-7-n+3n_1}{3}}
\label{55}
\end{equation}
The general solution of this equation is given by 
\begin{equation}
\rho_R=\beta\Big(\frac{n-2}{4-n-n_1}\Big)\rho_{BH0}\Big(1-\frac{3}{2}f\Big)^{-n+1}\alpha_0^{-3} \Big(1-\frac{t_c}{t}\Big)^{\frac{-1-n}{3}} t^{\frac{-4-n+3n_1}{3}}+Dt^{\frac{4n_1-8}{3}}
\label{56}
\end{equation}
where D is an integration constant determined from the initial value of radiation energy density. In this way we found the most general solution for the evolution of the black holes with matter and radiation in a matter-dominated universe.
In this solution for $\rho_R$, the second term gives the conventional evolution of radiation in a matter-dominated universe. But at the late times it is the first term which dominates.
Now by using equations (\ref{53}) and (\ref{56}) we can calculate the ratio of the energy densities of black holes and radiation as 
\begin{equation}
\frac{\rho_{BH}}{\rho_R}\to\frac{1}{\beta}\Big(\frac{4-n-n_1}{n-2}\Big)\Big(1-\frac{t_c}{t}\Big) 
\label{57}
\end{equation}
Here again we find for at late times $t\gg t_c$, equilibrium is attained. Thus accretion delays the onset of equilibrium. At equilibrium, the above ratio would be a constant depending only on the spectral index and $n_1$. Time dependence implies that the black holes and radiation tend towards an equilibrium where their energy densities scale the same way with time. One may also note that for $n>2$ equations (\ref{52}), (\ref{53}) and (\ref{56}) imply that black hole and radiation densities fall-off faster than matter justifying our earlier assumption. So our matter-dominated solution remains self-consistent. 

Proceeding in an analogous analysis as in case of radiation-dominated era one can easily establish the attractor property of the equilibrium solutions in matter-dominated era aswell.

We can also impose a constraint on the initial black hole density by comparing our theoretical expression with present observation. From observation of $\gamma$-ray background as well as those of the anti-protons from galactic sources impose bounds on the present day PBH density given by \cite{bijay1,bijay2,bijay3,bijay4,bijay5,bijay6,bijay7}
\begin{eqnarray}
{(\rho_{BH})_{t=t_0}} < 10^{-8} \rho_c
\label{biju1}
\end{eqnarray}
where $\rho_c$ is present density of the universe $\sim 1.1 \times 10^{-29} gm/cm^3$.

Again using different values of parameters of equations (\ref{53}), (\ref{54}) like $n_1\approx 0.00007$ \cite{nsmprd1,nsmprd2}, $n=3$ and $C \approx O(1)$, we found for absence of accretion $(f=0)$
\begin{eqnarray}
(\rho_{BH})_{t=t_0} \approx  3.35\times 10^{-5} \rho_{BH0}<10^{-37}gm/cm^3
\label{biju2}
\end{eqnarray}

Comparison of above two equations (\ref{biju1}) and (\ref{biju2}) gives
\begin{eqnarray} 
\rho_{BH0} < 3.28 \times 10^{-33} gm/cm^{3}  
\label{biju3}
\end{eqnarray}

The above equation gives the constraint on the initial PBH density. If we include accretion then the constraint will be multiplied by a term of $\Big(1-\frac{3}{2}f\Big)^{-n+1} $ which is typically of $O(10^{-1})$ with no significant effect on the bound given by equation (60). 

\section{Evolution of primordial black holes in an accelerated expanding universe}
The finding of SNIa observations that the universe is  currently undergoing accelerated expansion constitutes the most intriguing discovery in observational cosmology of recent years. As a possible theoretical explanation it is considered that the vacuum energy with negative pressure termed as dark energy is responsible for this acceleration. In this section our aim is to study evolution of primordial black hole mass spectrum in vacuum-dominated universe. For this we have to write Friedmann equation by introducing radiation, matter and vacuum energy terms.

Assuming a power law behaviour of $G(t) \sim \Big(\frac{1}{\phi(t)}\Big)$ in Friedmann equation for vacuum-dominated era i.e. $\frac{{\dot{\alpha}}^2}{{\alpha}^2} + \frac{\dot{\alpha}}{\alpha} \frac{\dot{\phi}}{\phi} -\frac{\omega}{6} \frac{{\dot{\phi}}^2}{\phi^2} =  \frac{8\pi}{3\phi} \rho_c$ and matching the time dimension of each term we found $G(t)$ varies with time like
\begin{equation}
G(t)= G_{0} \Big(\frac{t_0}{t}\Big)^{2}
\label{dwi1}
\end{equation}
Incorporating this time-dependence, the Friedmann equation becomes
\begin{eqnarray}
{\dot\alpha}^2 = \frac{8\pi {G_0}}{3}\Big(\frac{t_0}{t}\Big)^2 \alpha^2\Big[\rho_R +\rho_M + \rho_V + \rho_{BH0}\Big(1-\frac{3}{2} f\Big)^{-n+1} \alpha^{-3}(t)(t-t_c)^{\frac{2-n}{3}} \Big]\nonumber \\-\Big(\frac{2}{t}\Big)\dot{\alpha}\alpha +\Big(\frac{\omega}{6}\Big)\Big(\frac{4}{t^2}\Big)\alpha^2.
\label{dwi2}
\end{eqnarray}
Assuming $\rho_R$, $\rho_M$  and $\rho_{BH}$ are much smaller than $\rho_V$ in vacuum dominated era, we get
\begin{equation}
\frac{{\dot{\alpha}}^2}{\alpha^2} + \frac{2}{t}\frac{\dot\alpha}{\alpha}-\Big(\frac{8\pi {G_0}}{3}\frac{{t_0}^2}{t^2}\rho_V + \frac{2}{3}\frac{\omega}{t^2}\Big)= 0 
\label{dwi3}
\end{equation}
Solving this equation, we obtain
\begin{equation}
\alpha(t)= \alpha_{0}t^{\Big(-1+\sqrt{1+\frac{8\pi}{3}G_{0}t_{0}^2\rho_V+\frac{2}{3}\omega}\Big)} 
\label{dwi4}
\end{equation}
Using this value of $\alpha$(t) in equation (22), we can express black hole energy density as
\begin{equation}
\rho_{BH}(t)=\rho_{BH0}\alpha_{0}^{-3}\Big(1-\frac{3}{2}f\Big)^{-n+1}\Big(1-\frac{t_c}{t}\Big)^\frac{2-n}{3}t^\frac{2-n-9c_1}{3}
\label{dwi5}
\end{equation}
where $c_1 = -1+\sqrt{1+\frac{8\pi}{3} G_{0}t_{0}^{2}\rho_{V} + {\frac{2}{3}}\omega}$

Again using the value of $\alpha$(t) from equation (64), we can write the radiation energy equation (49) as 
\begin{equation}
\dot{\rho_R}= -\frac{4c_1}{t}\rho_R + \beta\Big(\frac{n-2}{3}\Big)\rho_{BH0}\Big(1-\frac{3}{2}f\Big)^{-n+1}\alpha_0^{-3}\Big(1-\frac{t_c}{t}\Big)^{\frac{-1-n}{3}}t^{\frac{-1-n-9c_1}{3}}
\label{dwi7}
\end{equation}
The general solution of this equation is
\begin{equation}
\rho_R=\beta\Big(\frac{n-2}{2-n+3c_1}\Big)\rho_{BH0}\Big(1-\frac{3}{2}f\Big)^{-n+1}\alpha_0^{-3}\Big(1-\frac{t_c}{t}\Big)^{\frac{-1-n}{3}}t^{\frac{2-n-9c_1}{3}} + Ft^{-4c_1}
\label{dwi8}
\end{equation}
where F is an integration constant determined from the initial value of radiation energy density.
In this way we found the most general solution for the evolution of black holes  with radiation, matter and vacuum energy in a vacuum dominated universe. In this solution for $\rho_R$, the second term gives the conventional evolution of radiation in a vacuum-dominated universe. But at the late time, it is the first term which dominates.

Now by using equations (\ref{dwi5}) and (\ref{dwi8}), we can calculate the ratio of the energy densities of black holes and radiation as

\begin{equation}
\frac{\rho_{BH}}{\rho_R}\to\frac{1}{\beta}\Big(\frac{3c_1+2-n}{n-2}\Big)\Big(1-\frac{t_c}{t}\Big)
\end{equation}
For late times $t\gg t_c $, here equilibrium is also attained and hence one can easily establish the attractor property of the equilibrium solutions in a vacuum dominated era by following the analysis of the previous sections.

We wish to mention in passion that no sensible constraint on $\rho_{BH0}$ can be obtained using our analysis, because $\rho_{BH}$(t) nearly vanishes for the vacuum dominated accelerated epoch as clear from Equation (\ref{dwi5}).


\section{Discussion and Conclusion}
In this paper we have studied the evolution of a power-law mass spectrum of PBHs by including accretion of radiation by PBHs along with Hawking evaporation within the context of Brans-Dicke theory. Here we also include the effects of the black hole evaporation on the expansion dynamics of the universe. We find that in radiation dominated era, there exists an equilibrium between energy densities of PBHs and radiation. In order to maintain the equilibrium, the PBHs are feeding a substantial amount of energy into the radiation via their evaporation.  Inclusion of accretion of radiation delays the onset of this equilibrium but once it is set up, accretion plays no further role. The positiveness of the ratio between energy densities of PBHs and radiation constrains the spectral index $n$ to lie between $2$ and $7/2$. But we find a stable exact solution in which the scale factor expands as $a(t) \propto t^{(8-n)/9}$. Thus we see that the scale factor evolves at a rate intermediate between that of a pure radiation-dominated (when $n \to 7/2$) and nearly matter-dominated (when $n \to 2$) universe. Then we study the nature of the stability of the equilibrium solution and get that the solution is asymptotically stable critical point of the full system and is a node. So for all physical initial conditions, the equilibrium solution act as an attractor. We have also shown similar results to hold good in matter and vacuum-dominated era.


\end{document}